\documentclass{epl}

\title{Bjerrum Pairing Correlations at Charged Interfaces}
\author{A. Travesset\inst{1} \and D. Vaknin \inst{1}}
\institute{
  Department of Physics and Astronomy and Ames Laboratory, Iowa State University, Ames, IA
  ,50011, USA }

\begin{document}

\maketitle

\begin{abstract}

Electrostatic correlations play a fundamental role in aqueous
solutions. In this Letter, we identify {\em transverse} and {\em
lateral} correlations as two mutually exclusive regimes. We show
that the transverse regime leads to binding by generalization of
Bjerrum pair formation theory, yielding binding constants from
first principle statistical mechanical calculations. We compare
our theoretical predictions with experiments on charged membranes
and Langmuir monolayers and find good agreements. We contrast our
approach with existing theories in the strong coupling limit and
on charged modulated interfaces, and discuss different scenarios
that lead to charge reversal and equal-sign attraction by
macro-ions.

\end{abstract}

\section{Introduction}

In recent years it has been recognized that the precise ion
distribution next to charged macromolecules is a key problem for
understanding biological processes such as cell signaling,
membrane fusion or DNA replication and is also of fundamental
relevance in many industrial applications. Ample theoretical and
experimental effort, reviewed in \cite{Bor:05,Gros:02}, have been
devoted to the problem.

The standard theoretical approach for describing ions in solution
next to charged interfaces is the Poisson-Boltzmann (PB) theory.
PB theory, however, ignores correlations, which are believed to
play a fundamental role when the interfaces are strongly charged
and the solution contains multi-valent ions. A new strong coupling
(SC) regime, defined by $\Gamma\equiv a_C/\lambda_G>> 1$, where
$a_C$ is a typical counterion separation near the interface and
$\lambda_G$ is the Gouy-Chapman length, has been identified, and
different SC theories have been proposed
\cite{Shkl:99,More:01,Bura:04,San:05}. These theoretical
descriptions usually assume that the charge at the interface is
smeared to a uniform density, whereas realistic interfaces consist
of discrete charges. Theoretical models incorporating the
discreteness of interfacial charges have been introduced recently,
both within SC \cite{MorNe:02,HSPP:04} and PB
\cite{LukSa:02,Trav:05}.

\begin{figure}
\begin{center}
\onefigure[height=1 in]{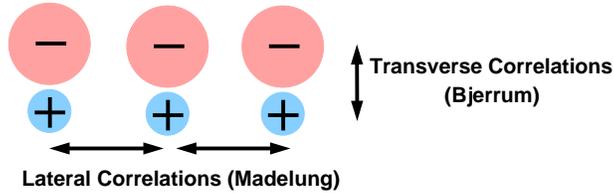}  \caption{Schematic view of the
transverse and lateral correlation regimes discussed in this
paper.}\label{f.0}
\end{center}
\end{figure}

Rather surprisingly, there are many experimental examples that
show that PB theory combined with a Langmuir adsorption theory
(LPB) \cite{Grah:48}, where ion adsorption to the interface is
empirically included by binding-constants, adequately describes
divalent ion distributions, both near membranes\cite{Macl:89} and
Langmuir monolayers \cite{Bloch:90}. It has been shown, for
example, that divalent ion distributions near fatty acid charged
Langmuir monolayers ($\Gamma\approx 20$) are described by LPB with
a striking accuracy \cite{Bloch:90}. This result is somewhat
unexpected, as it seems inescapable that tightly bound divalent
ions to the interface are laterally correlated, which should
result in an additional contribution to the LPB free energy (the
Madelung energy) which is given by (\cite{Trav:05a} and references
therein)
\begin{equation}\label{Madelung}
F_{corr}\approx -2.101 \sqrt{2} \frac{l_B}{a_L} k_B T \approx -4.2
k_B T
\end{equation}
where $l_B=7.1$\AA \ is the Bjerrum length and $a_L \approx
4.8$\AA \ is the lattice constant of a fatty acid in the
crystalline phase. As shown below (see fig.~\ref{f.4}), when the
Madelung contribution Eq.~\ref{Madelung} is accounted for within
LPB, the agreement between theory and experiment is completely
ruined. In this Letter, we consider {\em transverse} (as opposed
to {\em lateral}) correlations (see fig.~\ref{f.0}) between the
counterions and the discrete interfacial charges and show that
they induce binding by generalizing Bjerrum theory \cite{Rob:59}.
The free energy of the system becomes equivalent to LPB but now
the binding-constants are electrostatic in origin and can be
computed explicitly.

The Bjerrum pairing theory establishes that opposite charged ions
of valence $q_{+}$ and $q_{-}$ and radii $r_{+}$ and $r_{-}$ in
bulk solution form pairs, with an association constant $K_B$
(defined by $K_B=[AB]/[A^+][B^-]$, where $[X]$ is the electrolyte
concentration) given by \cite{Rob:59}
\begin{equation}\label{Bjerrum}
K_B = 4 \pi \int_{d}^{|q_{+} q_{-}| l_B/2} dr r^2 \exp(-q_{+}q_{-}
l_B/r) =4 \pi (|q_{+} q_{-}| l_B)^3 G(\frac{|q_{+} q_{-}|l_B}{d})
\end{equation}
where $d=r_{+}+r_{-}$ and $G(x)=\int^x_2 dz z^{-4} e^z$. We
interpret Bjerrum pairing as implying that if two oppositely
charged ions come closer than a distance $D \leq
\frac{|q_{+}q_{-}| e^2}{ 2 \varepsilon k_B T}\equiv
|q_{+}q_{-}|l_B/2$ they attract more strongly than the disordering
thermal fluctuations and bind. First principles expressions for
Bjerrum constants \cite{Petr:71} are numerically indistinguishable
from Eq.~\ref{Bjerrum}. Bjerrum theory has been recently
emphasized in many contexts such as, for example, in Coulomb
criticality \cite{Lev:96}.

\section{Correlations in the zero temperature limit}

\begin{figure}
\twofigures[height=1.7 in]{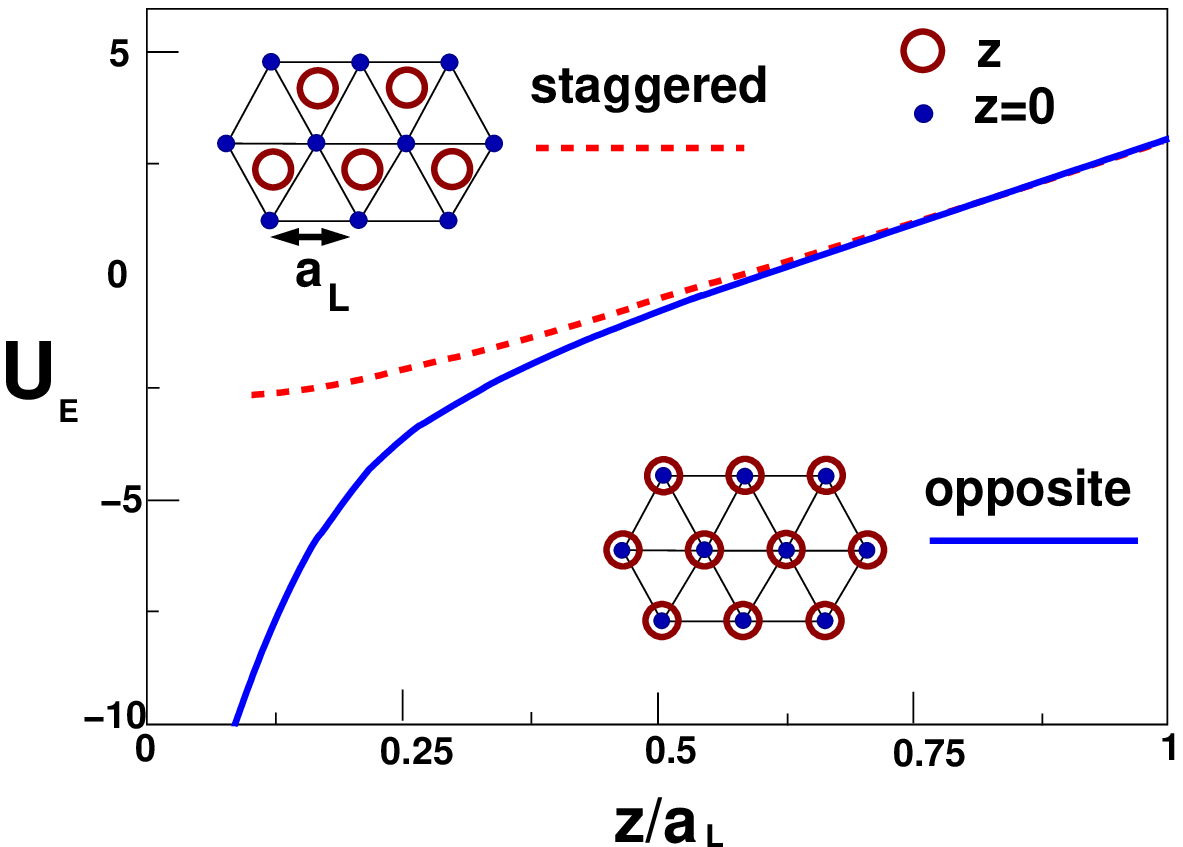}{Elec_Ratios.eps}
\caption{\label{f.1} Electrostatic energy $U_E$ (in units of $e^2
q^2/\varepsilon a_L$) as a function of separation for two
triangular two-dimensional lattices of opposite charge (of equal
valence $q$) in both the opposite and staggered configuration
}\caption{Electrostatic energy parameterized by the two ratios
defined in Eq.~(\ref{ratios}) defining the interface-ion and the
ion-ion regimes for the opposite configuration.\label{f.2}}
\end{figure}

In some electrostatic problems, insight into finite temperature
regimes can be obtained from the analysis of the zero temperature
limit \cite{Trav:05,Trav:05b,Trav:05a}. As a simple model for
quantitatively analyzing correlations, we consider the extreme
case of $N_P$ charged particles arranged as a two-dimensional (2D)
triangular crystal, interacting with $N_P$ oppositely charged
particles arranged as another 2D triangular crystal, see
fig.~\ref{f.1}. Both crystals have the same lattice constant $a_L$
and area of unit cell $A_C$. We compute the electrostatic energy
as a function of the distance $z$ between the two planes for two
situations, where the lattices are either opposite or staggered to
one another (see fig.~\ref{f.1}). This calculation is performed
exactly by Ewald summation techniques (see \cite{Trav:05a} and
references therein). Figure~\ref{f.1} shows that the opposite
configuration has lower electrostatic energy and we therefore
analyze it in more detail.

As a function of the separation ($z$), the electrostatic energy
can be divided into two regimes. The transverse correlation (TC)
regime ($z<<a_L$), where the energy is dominated by the attraction
of close opposite sign charges
\begin{equation}\label{Tot_energy_S}
U_S(z)=- N_P \frac{e^2}{\varepsilon z} \ ,
\end{equation}
and the lateral correlation (LC) regime ($z>>a_L$) where the
energy consists of the two Madelung energies $U_M$ (one for each
individual lattice) plus the energy of a planar capacitor with the
smeared surface charge of the lattices and width $z$. More
precisely
\begin{equation}\label{Tot_energy_L}
U_L(z)=N_P(2U_M + \frac{2 \pi e^2 z}{\varepsilon A_C}) .
\end{equation}

We characterize the two regimes (schematically shown in
fig.~\ref{f.0}) by defining two ratios measuring deviations of the
electrostatic energy $U_E(z)$ relative to their asymptotic values
Eqs.~\ref{Tot_energy_L} and \ref{Tot_energy_S}
\begin{equation}\label{ratios}
  r_1(z) = \frac{U_E(z)-2 N_P U_M}{\frac{2 \pi e^2 z N_P}{A_C}}  \ \ \ \ \ , \ \ \
  r_2(z) = \frac{U_E(z)}{U_S(z)}
\end{equation}
The transition separating the two regimes is sharp
(fig.~\ref{f.2}), defining a new characteristic length $z_0 \sim
0.35 a_L$ separating the TC and LC regimes. The length $z_0$ is a
property of the interface, independent of solution conditions
(ionic strength, temperature, etc.).

\section{Bjerrum pairing induced correlations}

We first discuss the simpler situation of a uniform charged
interface with counter-ions of valence $q$. The first equation of
the Ybon-Born-Green (YBG) hierarchy is an exact relation between
the ion distribution $n(z)$, and the counterion-counterion pair
distribution function $g({\bf r},{\bf r}^{\prime})$ \cite{Hans:02}
\begin{equation}\label{YBG}
    -k_B T \frac{dn(z)}{dz} = -q \frac{2 \pi \sigma}{\varepsilon} n(z)
    +\int d^{3}{\bf r}^{\prime} n(z) n({z}^{\prime})g({\bf r},{\bf
    r}^{\prime})\frac{\partial V(|{\bf r}-{\bf r}^{\prime}|)}{\partial z}
\end{equation}
where $V(r)= q^2 e^2/(\varepsilon  r)$. If $g({\bf r},{\bf
r}^{\prime})=1$ (no correlations) the above equation is equivalent
to PB. If, however, we assume a strong electrostatic repulsion
where each ion excludes every other like-sign ion $g({\bf r},{\bf
r}^{\prime})=0$ then the calculated density profile is
\begin{equation}\label{SC_density}
    n(z)=\frac{1}{2\pi l_B \lambda_G^2}
    \exp{(-z/\lambda_G)} \ ,
\end{equation}
where $\lambda_G$ is the Gouy-Chapman length. This result is the
counter-ion density profile within SC \cite{Shkl:99,More:01}. The
second equation in the YBG hierarchy can now be used to obtain the
correction to $g=0$, leading to $g({\bf r},{\bf
r}^{\prime})=\exp(-\frac{q^2l_B}{|{\bf r}-{\bf r}^{\prime}|}) \ $.
This result establishes the range of validity for the
approximation $g\approx 0$ as defined by the condition $|{\bf
r}-{\bf r}^{\prime}| << q^2 l_B$.

We now consider discrete interfacial charges. If $q_{-}$ is the
valence of the discrete interfacial charges, the same steps as
above give the counterion density $ n({\bf r}) \propto
\exp(\frac{|q_{+} q_{-}| l_B}{|{\bf r}-{\bf r}^{\prime}|})\ , $
which is asymptotically exact for $z<< |q_{+}q_{-}|l_B$. Mobile
ions ``see'' the interfacial charge as if no other charges are
present in the system. From the Bjerrum argument outlined in the
introduction, interface-ion pair-association then follows with a
binding constant
\begin{equation}\label{chem_const_bare}
K_I=\frac{K_B}{2}
\end{equation}
where $K_B$ has been defined in Eq.~\ref{Bjerrum}. The $1/2$
prefactor in Eq.~(\ref{chem_const_bare}) results from the fact
that half of the space (let us say the $z<0$ region) becomes
inaccessible to the mobile ions because of the presence of the
rigid interface. Selected values for the constants
Eq.~\ref{chem_const_bare} are given in table~\ref{t.1}.

\begin{table}
\begin{center}
\begin{tabular}{|c||c|c|c|c|c|c|c|c|}\hline
  Interface charge ($q_I$) & $Na^{+}$ & $K^{+}$  & $Cs^{+}$   & $Ca^{2+}/Cd^{2+}$ & $Ba^{2+}$ & $Mg^{2+}$ & $Mg^{2+}_H$ & $La^{3+}$ \\\hline
  $-1$     & $-0.44$ & $-0.63$ & $-1.05$   & $+0.98$  & $+0.90$  &   $1.10$      &   $0.23$  & $+1.869$  \\\hline
  $-2$     & $+1.00$ & $+0.91$ & $+0.82$   & $+2.80$  & $+2.53$  &   $3.22$      &   $1.80$   & $+4.965$  \\\hline
\end{tabular}
\caption{Values in $pK_I$ units ($pK_I=\log_{10}(K_I)$) for the
binding constants Eq.~(\ref{chem_const_bare}) of different
counterions to oxygen ($r_{-}=1.6$\AA). The bare
(crystallographic) radius $r_{+}$ of the ions is used except for
Mg$^{2+}_H$, for which the hydrated radius is used.(Units of $K_I$
are $1$M$^{-1}$).}\label{t.1}
\end{center}
\end{table}

\section{Free energy and counter-ion profiles}

The free energy of the system consists of a ``Stern'' layer of
bound ions induced by TC and a bulk solution containing the
remaining unbound ions. If $n_a^B$ is the bulk concentration of
counterions of type $a$, $f_a$ the fraction of bound ions and
$K^a_L$ the association constant of ion $a$ to the interfacial
charged groups, the free energy is
\begin{equation}\label{Free_Energy}
    \frac{F}{N_P k_B T}=-\sum_{a} f_a \ln(K^a_L n_a^B)+\left(1-\sum_{a} f_a\right)\ln\left(1-\sum_{a} f_a\right)+\sum_{a} f_a
    \ln(f_a)+\frac{F_{PB}(\sigma(f_a))}{N_P k_B T}.
\end{equation}
The binding free energy gain for $f_a N_P$ ions is $-f_a N_p
\ln(K^a_L)$ and the loss of entropy is $-N_P f_a \ln(n_a^B)$.
These two contributions combine in the first term of the free
energy. The next term is the mixing entropy of the interfacial
species, and the last term $F_{PB}$ is the free energy of the
remaining unbound ions. We restrict the analysis to the dilute
limit, where the activity coefficient of the ions may be
approximated as unity, so $F_{PB}$ is the PB free energy of a
solution containing the remaining free (not bound) ions, thus
keeping the number of free and bound ions constant. Minimization
of the free energy with respect to $f_a$ leads to
\begin{equation}\label{Langm}
    f_a=\frac{K^a_L n_a^B \exp(-q_a \phi(0)) }{1+\sum_{a}K^a_L n_a^B \exp(-q_a
        \phi(0))}.
\end{equation}
The effective surface charge density is $
    \sigma(f_a)=\sigma_0 \frac{-1+\sum_{a}(q_a-1)K^a_L n_a^B
    \exp(-q_a \phi(0))}{1+\sum_{a}K^a_L n_a^B \exp(-q_a
    \phi(0))} \ $,
where $\sigma_0=-e/A_c$ and $\phi(0)$ is the contact value
potential in units of $k_B T/e$. It is assumed that counterions
bind in a 1:1 ratio to the charges at the interface. It is trivial
to include binding in a 2:1 (2 surface charges to 1 counterion)
ratio, which may occur in some membranes \cite{Hust:01}.

From our previous discussion it should be expected that $K^a_L$ is
given by $K_I^a$, Eq.~\ref{chem_const_bare}. It should be noted,
however, that $K^a_I$ is obtained by integration of a half-sphere
of radius $|q_I q_{a}|l_B/2$, where the mobile ion ``sees'' the
interfacial charge as the only charge in the system. From the
discussion following fig.~\ref{f.2}, however, the mobile ion can
only ``see'' the interface charge if it is within a distance
$z_0\sim 0.35 a_L$ from the interface. There are therefore two
cases that need to be considered. If $z_0 > |q_I q_{a}|l_B/2$ the
association constant is given by Eq.~\ref{chem_const_bare}, but if
$z_0 \ll |q_I q_{a}|l_B/2$ the integration defining
Eq.~\ref{chem_const_bare} is restricted to a much smaller domain
and the actual constant $K_L$ is smaller than $K_I$,
\begin{equation}\label{Act_binding}
     K^a_L=K^a_I \ \ ( a_L \geq \alpha |q_{I} q_{a}| l_B ) \ \ \ \ , \ \ \ \  K^a_L \ll K^a_I \ \ ( a_L \ll \alpha |q_{I} q_{a}| l_B
     )\ ,
\end{equation}
where $\alpha\approx 1$--$1.4 $. In cases where the ion binding is
covalent, the previous formulas do not apply. An important example
is the proton, but the association constants for the proton follow
from pK$_{a}$ values ( pK$_a$=2.16 and pK$_a$=5.1 for phosphate
and carboxyl groups, respectively).

\section{Comparison with experiment}

\begin{table}
\begin{center}
\begin{tabular}{|c||c|c|c||c|c|c||c|c|c|}\hline
  \multicolumn{1}{|c||}{}& \multicolumn{3}{c||}{pK$_I$
  (membrane)}& \multicolumn{6}{c|}{pK$_B$(
  solution)}\\\hline
  Ion          & Eq.~\ref{chem_const_bare} & Exp.~\cite{McLa:81} & Exp.~\cite{Hust:01} & Eq.~\ref{Bjerrum}(-1)  & Carb.  & Phos. & Eq.~\ref{Bjerrum}(-2) & Carb. & Phos. \\\hline
  $Na^{+}$     & $-0.44$ & $-0.6$         & $-0.35$        &      $-0.14$             & $$      & $$  &       $0.90$      &   $$  & $$  \\\hline
  $Ba^{2+}$    & $+0.90$ & $$             & $$             &      $+1.2$              & $$      & $$  &       $2.83$      &   $2.83$   & $$  \\\hline
  $Ca^{2+}$    & $+0.98$ & $1.0$          & $$             &      $+1.3$              & $1.0$   & $1.4$  &    $3.1$      &   $3.15$   & $2.74$  \\\hline
\end{tabular}
\caption{Comparison between experimental and theoretical
(Eq.~\ref{chem_const_bare} and Eq.~\ref{Bjerrum}) pK values.
Carbonic (Carb.) and Phosphoric (Phos.) pK values are reported in
\cite{CSC} for single (-1) and double (-2) deprotonated
acids.}\label{t.2}
\end{center}
\end{table}

\begin{figure}
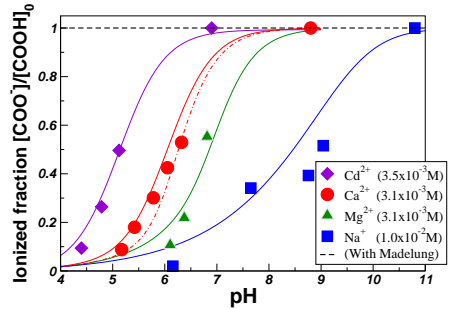

\begin{center}
 \onefigure[height=1.6 in]{dissociation.eps}
\caption{Dissociation of arachidic
 acid as a function of pH. The symbols are the experimental results from
 \cite{LeCal:01}. The solid lines are the results from Eq.~\ref{Free_Energy} with
 a 1:1 binding. The dashed-dotted line corresponds to a 2:1 binding for calcium and the
 dashed line shows the result of including the Madelung term Eq.~\ref{Madelung}
 into the free energy Eq.~\ref{Free_Energy}.}\label{f.4}
 \end{center}
\end{figure}

We selected three ions with the atomic structure of the noble gas:
Na$^{+}$,Ba$^{2+}$ and Ca$^{2+}$, binding to carboxyl (-COOH) and
phosphate (-PO$_4$H) groups (in both cases, the binding is to an
oxygen atom) and compared our predictions to available
experimental results in table.~\ref{t.2}. In order to assess the
validity of using the bare radius for oxygen in
Eq.~\ref{chem_const_bare}, we included binding constants to small
soluble molecules containing the same groups, which are described
by Eq.~\ref{Bjerrum}. The agreement with experiment is
satisfactory.

Fatty acids are strongly charged systems ($\Gamma\approx 20$) and
are appropriate for investigations of the SC limit. A first
qualitative prediction of our approach is that divalent ions
should form a tight "Stern" layer, with almost negligible
distribution, next to the interface. This is well established
experimentally from many X-ray reflectivity studies after the
pioneering work by Kjaer et al. \cite{Kjaer:89}. On a more
quantitative level, we use the data on arachidic acid (AA)
obtained by infrared reflection-absorption spectroscopy
(IRRAS)\cite{LeCal:01}.

Figure~\ref{f.4} shows IRRAS experimental results \cite{LeCal:01}.
The solid line plots the degree of dissociation expected from the
LPB free energy Eq.~\ref{Free_Energy}. We used the expected
pK$_{a}=5.1$ of the carboxyl group and fitted the association
constant to the experimental data. For calcium, pK $=0$ is
obtained. This value is one order of magnitude smaller than the
predicted value in Table.~\ref{t.1}, but this should be expected
given Eq.~\ref{Langm} and the fact that typical interfacial charge
separation (the lattice constant for AA) satisfies $a_L\approx
4.8$\AA$<< 2 l_B = 14.2$\AA. For Cd$^{2+}$ pK $=1.8$ is obtained.
This value seems in contradiction with table~\ref{t.2} and
Eq.~\ref{Langm}. However, Cd$^{2+}$ ions do not have a noble gas
structure and covalent bonding with oxygen is possible. This has
been confirmed from IRRAS data \cite{Kut:96}. We should note that
if the binding is indeed covalent, the pK value obtained should be
identical with the critical stability constants \cite{CSC}
pK$=1.8$, in agreement with our result. The constant for Mg$^{2+}$
is pK$\approx-1.4$. Given that Mg$^{2+}$ ion has a smaller radius
than Ca$^{2+}$, Eq.~\ref{chem_const_bare} predicts a stronger
Mg$^{2+}$ binding than Ca$^{2+}$. In \cite{LeCal:01} it is argued
that Mg$^{2+}$ may bind covalently with AA, but by analogy with
the Cd$^{2+}$ case, we should expect an enhanced value for the
binding constant as compared with Ca$^{2+}$. It is well known that
Mg$^2+$ has a very stable hydration sheath \cite{Israe:00} and we
suggest that Mg$^{2+}$ binds electrostatically to oxygen retaining
its hydration sheath, (with a hydrated radius $r_{+}=4.3$ \AA
\cite{Israe:00}). When the hydrated radius is used in
Eq.~\ref{chem_const_bare}  (see Table.~\ref{t.1}), the calculated
association constant is consistent with the experimental values.
For Na$^{+}$ ions we find no binding (pK$<-2.5$). In general, we
infer that the small lattice constant of AA results in a reduction
of the binding constants by an order of magnitude or more from
those obtained by Eq.~\ref{chem_const_bare}. This is in agreement
with our own reflectivity data for $Cs^{+}$ ions ($r_{+}=1.7$\AA)
next to phosphate groups \cite{Bu:05} $a_L=6.8$\AA. We also
analyzed older experimental data \cite{Bloch:90} and found
complete consistency with the results. We were not able to find
similar data for trivalent ions.

\section{Conclusions}

In this paper we have identified two correlation regimes, namely
the TC and LC (fig.~\ref{f.0}). The TC regime results in
electrostatic binding to the interface, for which the association
constants are computed by generalizing the Bjerrum theory
\cite{Rob:59}. Our approach accounts for ion specificity (see
table.~\ref{t.1}) by including the finite size of the ions and the
nature of the head group charge through its size and from its
pK$_a$ value, which accounts for proton transfer and release. We
compared our theoretical calculations with different experimental
results and found good agreement, see table~\ref{t.2} and
fig.~\ref{f.4}.

The theory presented differs from previous theories
\cite{Shkl:99,More:01,Bura:04,San:05} in that these theories
assume a smeared interfacial charge distribution, where TC is
absent, and deal with the LC. A scenario for the LC would be, for
example, when interfacial charges are buried a distance
$d>0.35a_L$ inside the interface. Our approach differs from
previous theories on charge modulations \cite{MorNe:02,HSPP:04} in
that it incorporates binding by Bjerrum pairing into LPB. For
slightly charged modulated interfaces (defined by a small contact
value potential relative to $k_B T$) at low monovalent salt
concentrations PB including modulations \cite{LukSa:02,Trav:05}
still applies.

In some situations (defined by amphiphile geometry and valence) we
speculate that ions may penetrate inside the head group in a
staggered configuration shown in fig.~\ref{f.1}, forming a
``molten salt'' state consisting of the charged head groups and
the mobile ions. This ``molten salt'' state has a Madelung energy
(given by the limit $z\rightarrow 0$ in the corresponding
fig.~\ref{f.1}), and may lead to charge reversal. For moderate
salt concentrations (defined by a Debye length smaller than $a_L$,
the typical separation of charged interfacial groups), charge
reversal of the interface by binding is also possible. Charge
reversal has been experimentally observed in both membranes
\cite{Macl:89} and monolayers \cite{Vak:03}. We have not discussed
in any detail geometries other than the plane. We expect our
arguments on Bjerrum induced correlation binding to apply for
other geometries, such as cylinders, where equal charge attraction
follows from binding \cite{Are:99}.

It is imperative to compare theoretical ion distributions to
experimental distributions that include points distant from the
interface. Only recently, however, with the use of anomalous X-ray
reflectivity techniques, ion distributions are becoming available
\cite{Vak:03,Bu:05}. We hope to provide more detailed comparisons
with those experimental results in the near future.

\acknowledgments

The work of AT has been supported by NSF grant DMR-0426597. The
work at the Ames Laboratory is supported by the DOE, office of
Basic Energy Sciences under contract No. W-7405-ENG-82.


\begin{thebibliography}{0}

\bibitem{Bor:05}
  \Name{Boroudjerdi H. et al.}
  \REVIEW{Phys. Reports}{416}{2005}{129}.

\bibitem{Gros:02}
  \Name{Grosberg A.Yu., Nguyen T.T and Shklovskii B.I.}
  \REVIEW{Rev. Mod. Phys.}{74}{2002}{329}.

\bibitem{Shkl:99}
\Name{Shklovskii B.I.}
  \REVIEW{Phys. Rev. E}{60}{1999}{5802}.

\bibitem{More:01}
\Name{Moreira A.G., Netz R.}
  \REVIEW{Europhys. Lett.}{52}{2001}{705}.

\bibitem{Bura:04}
\Name{Burak Y.,Andelman D. and Orland H.}
  \REVIEW{Phys. Rev. E}{70}{2004}{016102}.

\bibitem{San:05}
\Name{Santangelo C.} \REVIEW{arXiv:cond-mat 0509007}{}{2005}{}.

\bibitem{MorNe:02}
\Name{Moreira A.G. and R.R. Netz} \REVIEW{Europhys. Lett.}
{57}{2002}{911}.

\bibitem{HSPP:04}
\Name{Henle M.L., Santangelo C.D., Patel D.M. and Pincus P.}
\REVIEW{Europhys. Lett.}{66}{2004}{286}.

\bibitem{LukSa:02}
\Name{Lukatsky D.B. and Safran S.A.}\REVIEW {Europhys. Lett.}
{60}{2002}{629}.

\bibitem{Trav:05}
\Name{Travesset A.} \REVIEW{Eur. Phys. J. E} {17}{2005}{435}.

\bibitem{Grah:48}
\Name{Grahame, D.C.} \REVIEW{Chem. Rev.}{1}{1947}{103}.

\bibitem{Macl:89}
  \Name{McLaughlin S.}
  \REVIEW{Annu. Rev. Biophys. Chem.}{18}{1989}{113}.

\bibitem{Bloch:90}
\Name{Bloch J.M. and Yun W.}
  \REVIEW{Phys. Rev. A}{41}{1990}{844}.

\bibitem{Trav:05a}
\Name{Bowick M. et al.} \REVIEW{Phys. Rev. B}{73}{2006}{24115}.

\bibitem{Rob:59}
  \Name{Robinson R.A. and Stokes R.H.}
  \Book{Electrolyte Solutions}
  \Publ{Dover Pub., Mineola, NY}
  \Year{1959}.

\bibitem{Petr:71}
  \Name{Petrucci, S.}
  \Book{Ionic Interactions}
  \Publ{Academic Press, New York, NY}
  \Year{1971}.


\bibitem{Lev:96}
\Name{Levin Y. and Fisher M.}
  \REVIEW{Physica A}{225}{1996}{164}.

\bibitem{Trav:05b}
\Name{Travesset A.} \REVIEW{Phys. Rev. E}{72}{2005}{36110}.

\bibitem{Hans:02}
  \Name{Hansen J.P. and McDonald I.R.}
  \Book{Theory of Simple Liquids}
  \Publ{Academic Press, London}
  \Year{2003}.

\bibitem{McLa:81}
\Name{McLaughlin S. and Brown J.}
\REVIEW{J. Gen. Physiol.}{77}{1981}{445}

\bibitem{Hust:01}
\Name{Huster D.,Arnold K. and Gawrisch K.} \REVIEW{Biophys.
J}{78}{2000}{3011}.

\bibitem{CSC}
  \Editor{Martell A.E. and Smith R.M.}
  \Book{Critical Stability Constants}
  \Publ{Plenum, NY}
  \Year{1974}.


\bibitem{Kjaer:89}
\Name{Kjaer et al.}\REVIEW{J. Phys. Chem.}{93}{1989}{3200}.

\bibitem{LeCal:01}
\Name{Le Calvez et al.} \REVIEW{Langmuir} {17}{2001}{670}.

\bibitem{Kut:96}
\Name{Simon-Kutscher J.,Gericke A. and Huhnerfuss H.}
\REVIEW{Langmuir}{12}{1996}{1027}.

\bibitem{Israe:00}
  \Name{Israelachvili J.}
  \Book{Intermolecular and surface forces}
  \Publ{Academic Press, London}
  \Year{2000}.

\bibitem{Bu:05}
\Name{Bu W.,Vaknin D. and Travesset A.} \REVIEW{Phys. Rev.
E}{72}{2005}{60501}.

\bibitem{Vak:03}
\Name{Vaknin D., Kruger P. and Losche M.} \REVIEW{Phys. Rev.
Lett.}{90}{2003}{178102}.

\bibitem{Are:99}
  \Name{Arenzon JJ.,Stilck J. and Levin Y.}
  \REVIEW{Eur. Phys. J. B}{12}{1999}{79}.

\end{thebibliography}
\end{document}